\documentstyle[11pt]{article}

\begin{document}

\title{ \bf Can the Equivalence Principle Survive
Quantization?\thanks{Plenary lecture at the International Workshop on
AntiMatter Gravity and AntiHydrogen Spectroscopy, Molise, Italy May 96
(to be published in the proceedings).} }

\author{ A.Y. Shiekh \\
International Centre for Theoretical
Physics, Miramare, Trieste, Italy }

\date{}

\maketitle

\begin{abstract}
It is well known that Einstein gravity is
non-renormalizable; however a generalized approach is proposed
that leads to Einstein gravity {\it after} renormalization.
This them implies that at least one candidate for quantum
gravity treats all matter on an equal footing with regard to
the gravitational behaviour.
\end{abstract}

\begin{quote}
{\small{\em ``It stands to the everlasting credit of
science that by acting on the human mind it has overcome man's
insecurity before himself and before nature.''}
-- Albert Einstein}
\end{quote}

\baselineskip = 18pt

\section{Introduction}
\subsection{Is Einstein gravity incomplete?}
It is intrinsic from the start that in Einstein gravity all
matter (anti-matter included) responds equally to the
gravitational field. However, Einstein gravity cannot be
quantized [reviewed in Isham, 1981], and as a result one might
wonder if it was the equivalence principle that was hindering
quantization. Nevertheless, the fact that it seems impossible to
quantize, does not exclude the existence of a quantum form of
Einstein gravity; in the same way as prehistoric man had a
source of fire (lightning strikes on trees), though not
necessarily a means to produce it [Shiekh, 94; 96]. It is all too easy to
forget that the world is quantum, and that the classical
picture is but a special case, so we are actually working
backwards when we try to derive the more general case from a
restrictive form, and there is no reason to believe that this
is always possible. However, we are compelled to follow this route
out of a lack of choice.

Even if there were to exist a quantum form of Einstein
gravity, this would prove very little except to present an
example of a complete theory of gravity that made a solid
prediction, namely that matter and anti-matter respond equally
to the gravitational field. It would reduce the compulsion,
though not the need, to ask nature her opinion.

There is also the line of reasoning that since the equivalence
principle is a founding part of Einstein gravity, it is biased
in this matter. But it is just such inflexibility that makes a
theory predictive, and without this behaviour a theory is only
descriptive. We should not forget that
we are happy to build in other factors such as the
conservation of momentum and energy into most of our theories.
Physics is not so much explaining things per se, but rather
explaining things in terms of the least number of reasonable
unknowns (reductionism).
One might still ask why momentum and energy are conserved,
and can partly answer this through Noether's theorem, which
tells us it follows from space and time invariance (the
physics of here and now is the same as that of there
and yesterday, or that of tomorrow). But at some point the
basic question remains open; for example, nobody really knows the
mechanism behind the instantaneous collapse of the wave
function, or the reality behind the infinities we turn back
upon the construct in the case of quantum field theory.
At some point we fall away
from hard science and are left with a few dangling philosophical questions.
However, there is progress in this reductionism, and we
should have a certain reluctance to adopting a less
predictive theory, unless compelled to do so by experiment.
Einstein gravity is based upon the aesthetically
pleasing picture of free flight in a curved space-time, where the
forces of Newton's picture are gone. Why or how it is curved is
not known, only by how much and by what.
Perhaps one day the back tracking will end, and all will be
explained in the maybe unsatisfying response that the
world is the way it is on the grounds that there is
only one self-consistent way to formulate it.

If we found ourselves unable to build a
{\it complete} theory without letting go of the underlying
principle, we would have very good reason to consider
abandoning it; and so, the existence of a candidate is
thought to have some significance, in having something to take to
the `lab'. One might even take the stance that a proposal amplifies
the need to test the validity of the underlying principles.

It is this approach I wish to take, with neither two open a
mind, nor one too closed. It would be wrong to so quickly
abandon all the efforts of the past, just as it would be
dangerous to adhere to them with excessive rigidity. And so
this will not be a revolutionary, scorched earth, proposal;
but a more evolutionary move. But we should not rush ahead,
for the proposal might be made that gravity need not be
quantized at all, that the classical theory is the
complete theory.

\subsection{Need gravity be quantize at all?}
Is it not possible that although the other forces of nature
are seen to be quantized, that perhaps gravity, which is
presently not seen as a force at all, need not be quantized
[Feynman, 1963; Kibble, 1981]? The concept of force is a
classical one that does not even make an appearance in quantum
theory, where distribution is governed by the overlap of wave
functions. Newtonian
gravity certainly carries the notion of force, while the Einstein
view is of matter in free fall in a curved space-time. This
picture of gravity might lead one to propose leaving the
curved space time unquantized, and to have the quantum fields
play out on this arena. It is an appealing scenario, and would
present a resolution to the problem of quantization, claiming
it is not necessary at all.

However the gravitational field, if left classical, could be used
to make measurements on a quantum field, and being
classical would not be subject to the Heisenberg uncertainty
principle. Thus if one were to look at a quantum field using
gravity as a probe, one would be able to extract information
about the quantum field that defied the Heisenberg principle.
Such arguments are far from water-tight, but do make a strong
case for the quantization of gravity.

Despite the very geometric picture that is usually assigned
to Einstein gravity, it can be made to look like the other
perturbative field theories, with the graviton a spin two
particle. It is this very traditional, particle physics
perspective that we will be following. Unfortunately, the whole
structure now takes on a very mathematical structure, and at
times it is hard to keep contact with the physics.

Actually, we will be adding a second theme to the quantization
that is not strictly necessary, namely renormalization by
analytic continuation. In these methods there is no subtraction
of infinities, but rather a reinterpretation of the formulae.
This is a formulation devoid of
infinities, so completely by-passing the problem of interpreting
the divergences. At the risk of mixing concepts, the
quantization is illustrated using this novel method, which can be
shown to be equivalent, in effect, to the older techniques.

\subsection{Tweaking gravity}
Einstein gravity is sometimes rather interesting in its
intolerance to tweaking. Normally one might suppose that one
is always free to add a small modification to a theory, so
long as it does not imply an intrinsic inconsistency. One
usually has the impression that experiment can never do more
than put limits on a small deviation. For example, the photon in electrodynamics
could be assigned a small rest-mass, so small that the effect
goes unnoticed. It is worth noting that this freedom is not
always present. One might suggest a very small rest
mass for the graviton; but in general relativity, the mass
goes to zero limit gives a differing predictions from the
massless theory [van Nieuwenhuizen, 1973], and for this reason alone,
we will not be employing a rest mass for the graviton to tame
the infra-red divergences that are encountered.

\subsection{Can anti-matter fall upwards?}
We are used to thinking of matter and anti-matter as
complementary. That is to say the charges of a
particle and it's anti-partner are normally either the same or
opposite. This opposition could not be the case for gravity, for I
recall an accelerator engineer saying that
gravity must be taken into account when designing todays particle
accelerators. This being so, a direct anti-gravity response of
anti-matter would have been observed long ago.
So any effect, if any, would be expected not to display the
anti-symmetry we are biased to think of, and would need to be
at a smaller level, perhaps even only as a quantum correction.
Also, this effect must some how not be present in the
particles we think of as being their own anti-particles (the
photon, for instance), unless one is willing to modify not
just gravity, but the entire spectrum of forces (particles).

\subsection{Should we look?}
No matter how beautifully the gears of a theory turn, it is
little more than a piece of art till it has gone out on the
field and proved its worthiness. After all, the null results
of Michelson and Morley supported special relativity, while
the null result of E\"{o}tv\"{o}s and Dicke lent support
to general relativity. So a null result of the violation on the
equivalence principle might be viewed as significant support
for quantum Einstein gravity.

Any non-zero deviation would make a turmoil of the usual
matter/anti-matter symmetry. Of course, one could propose
there are two types of mass, one (like the photon) that is its
own anti-particle, and so does not change sign; and the other
which does. The small deviations would then be explained as a
dominance of the first type. But this approach lacks
predictive power, with a tuning parameter to fit
experimental results.

\subsection{The problem with traditional quantization
(a lightning review)}

Those that know the traditional methods and problems with
quantizing are not in need of an introductory review,
while those that have not tinkered with the guts cannot really be
properly shown the approach in an hour. It is with this
contradiction in mind that we set out on a lightning tour of the
problem.

The normal approach is to start with a classical theory and
try to quantize it. In reality the world is quantum, and the
classical view is just a special limit. In this sense we are
starting from the top and trying to work down to the more
fundamental. It is an approach fraught with dangers, but it is
the best we can do for now. There is no guarantee that the
attempt to derive the more general from the special case will
bear fruit, and the fact that Einstein gravity can't be
quantized should not be taken to imply that there is no
quantum Einstein gravity.

We have been quantizing non-relativistic systems with success
for some time now, but relativistic theories seem to demand
the use of field theories (to allow for particle creation and
destruction). However, the infinite number of degrees of
freedom tends to be accompanied by infinite quantities in the
theory, and this, very crudely, is the source of the problem
when quantizing field theories. However, some field theories
are quantizable, despite the presence of these infinities. It
turns out that in some very special cases it is possible to
re-absorb the infinities into the coupling constants of the
original, starting (classical) theory. It is a mathematical
technique that is difficult to interpret physically; but
despite this difficulty it leads to very good physical
predictions for most of the forces of nature (the electric,
weak and strong forces). The fact that this so called process
of renormalization is only successful for as small class of
theories is what makes it predictive. In fact, it is so
successful that it has permitted the unification of the
electric and weak forces and even has a lesser constrained proposal
for uniting also the strong nuclear force. However, the one
remaining force, namely gravitation, does not succumb to
quantization so easily.

So long as this is the
case, speculation about the quantum regime has a free rein and
can run unhindered, and unguided. A convincing
theory should make hard predictions and act as a guide to what
should be tested by experiment. The more a
theory gambles in taking a hard stance, and risking being shown
wrong, the more it stands to gain should it be found to stand
in harmony with nature. The first World War saved Einstein
some embarrassment in this respect, for he was able to locate
a factor of two error in his calculation for the bending of
light, before Eddington was able to confirm
his prediction. A correction after the fact would have been a
lot less convincing.

The traditional methods of quantization that have worked so
very well in taming the theories of the other forces of nature,
have not fared so well when taking on the task of quantizing
gravity. Here the infinities of quantum field theory don't
match
the original coupling constants. This suggests the need for
something novel, but should not be taken as reason to totally
abandon the past thinking as a complete failure, and so devoid of
usefulness. It is a habit most every generation makes politically,
and this tends to hinder progress.

Despite this need for change, there seems to be a proposal for
quantizing gravity that is unexpectedly conservative in its lack
novelty. In fact, we will be so traditional as to investigate
the perturbative quantization of gravity. This means that much
of the presentation can be versed in the now rather old
language of field theory, and we can embark on a more
detailed investigation of the problems obstructing the
traditional quantization of Einstein gravity.

\section{The Perturbative Quantization of Einstein Gravity}

The usual scheme of field quantization is plagued by
divergences, but in some special cases those infinities can be
consistently ploughed back into the theory to yield a finite
end result with a small number of arbitrary constants
remaining; these then being obtained from experiment [Ramond,
1990; Collins, 1984]. This is the renown scheme of
renormalization, disapproved of by some, but reasonably well
defined and yielding results in excellent agreement with
nature. For those disturbed by the appearance of infinities,
there now exists a finite perturbative version employing
analytic continuation (a generalization of the Zeta function,
one loop, technique), that goes under the deceptive name of
`operator regularization'. The fact that after renormalization
some factors, such as mass and charge, are left undetermined
should perhaps not be viewed as a predictive shortcoming,
since the fundamental units of nature are relative. That is to
say, the choice of reference unit (be it mass, length, time,
or charge) is always arbitrary, and then everything else can
be stated in terms of these units. In this sense the final
theory of everything should not, and cannot, predict all.

The fact that only some theories are renormalizable has the
beneficial effect of being selective, and so predictive. This
follows the line of reasoning that is more than
{\it descriptive}, but {\it predictive} by virtue of being
limited by the requirement of self consistency.

Unfortunately, in the usual sense, general relativity is {\it
not} renormalizable [Veltman, 1976], and we will run quickly
over the failure of Einstein gravity to quantize
perturbatively, by considering the example of a massive scalar
field with gravity. The starting theory in Euclidean space
would be characterized by:

\begin{equation} L = -\sqrt{g}
\left( R +
\textstyle{1 \over 2} g^{\mu \nu}
\partial_\mu \phi \partial_\nu \phi +
\textstyle{1 \over 2} m^2 \phi^2
\right)
\end{equation}

\rightline{ \small \it
(using units where $16 \pi G = 1$, $c = 1$)}

One discovers, upon perturbatively quantizing both the matter
and gravitational fields, that the counter terms do not fall
back within the original Lagrangian, so the infinities cannot
be reabsorbed.

One natural thought might be to generalize Einstein
gravity by extending the starting Lagrangian to accommodate the
anticipated counter terms. Here symmetry can be employed, and
by using the most general starting Lagrangian consistent with
the original symmetries one arranges that the counter terms
(which also retain the symmetry in the absence of an anomaly)
fall back within the Lagrangian. One would not anticipate an
anomaly, as these arise from a quantum conflict between two or
more symmetries, when one must choose between one or the
other. Thus one is lead to the infinitely large Lagrangian:

\begin{equation} L_0 = -\sqrt{g_0}
\left(\matrix { -2\Lambda_0 + R_0 +
\textstyle{1 \over 2} p_0^2 +
\textstyle{1 \over 2} m_0^2 \phi_0^2 +
\textstyle{1 \over 4!} \phi_0^4
\lambda_0(\phi_0^2) + p_0^2 \phi_0^2
\kappa_0(\phi_0^2) + R_0 \phi_0^2 \gamma_0(\phi_0^2) \cr
\cr + p_0^4 a_0 (p_0^2,\phi_0^2) + R_0 p_0^2
b_0(p_0^2,\phi_0^2) + R_0^2 c_0(p_0^2,\phi_0^2) + R_{0\mu\nu}
R_0^{\mu\nu} d_0(p_0^2,\phi_0^2) + ... }
\right)
\end{equation}

\noindent where $p_0^2$ is shorthand for $g_0^{\mu \nu}
\partial_\mu \phi_0 \partial_\nu \phi_0$ and not the
independent variable of Hamiltonian mechanics.
$\lambda_0$, $\kappa_0$, $\gamma_0$, $a_0$,
$b_0$, $c_0$, $d_0$ ... are arbitrary analytic functions, and
the second line carries all the higher derivative terms.
Strictly this is formal in having neglected gauge fixing and
the resulting presence of ghost particles.

The price for having achieved `formal' renormalization, is
that the theory (with its infinite number of arbitrary
renormalized parameters) is devoid of predictive content. The
failure to quantize has been rephrased from a
problem of non-renormalizability to one of non-predictability.

Despite this, after renormalization we are lead to:

\begin{equation} L = -\sqrt{g}
\left(\matrix { -2\Lambda + R +
\textstyle{1 \over 2} p^2 +
\textstyle{1 \over 2} m^2 \phi^2 +
\textstyle{1 \over 4!} \phi^4 \lambda(\phi^2) + p^2 \phi^2
\kappa(\phi^2) + R \phi^2 \gamma(\phi^2) \cr \cr + p^4 a
(p^2,\phi^2) + R p^2 b(p^2,\phi^2) + R^2 c(p^2,\phi^2) +
R_{\mu\nu} R^{\mu\nu} d(p^2,\phi^2) + ... }
\right)
\end{equation}

However, there remain physical criterion to pin down some of
these arbitrary factors. Since in general the higher
derivative terms lead to acausal classical behavior, their
renormalized coefficient can be put down to zero on physical
grounds. This still leaves the three arbitrary functions:
$\lambda(\phi^2)$, $\kappa(\phi^2)$ and
$\gamma(\phi^2)$, associated with the terms $\phi^4$,
$p^2 \phi^2$, and
$R \phi^2$ respectively. The last may be abandoned on the
grounds of defying the equivalence principle. To see this,
begin by considering the first term of the Taylor expansion,
namely $R\phi^2$; this has the form of a mass term and so one
would be able to make local measurements of mass to determine
the curvature, and so contradict the equivalence principle
(charged particles, with their non-local fields have this term
present with a fixed coefficient). The same line of reasoning
applies to the remaining terms, $R\phi^4$, $R\phi^6$, ... etc.

This leaves us the two remaining infinite families of
ambiguities with the terms $\phi^4\lambda(\phi^2)$ and
$p^2\phi^2\kappa(\phi^2)$. In the limit of flat space in 3+1
dimensions this will reduce to a renormalized theory in the
traditional sense if $\lambda(\phi^2)=constant$, and
$\kappa(\phi^2)=0$. So one is lead to proposing that the
physical parameters should be:

\begin{equation}
\matrix {
\Lambda = \kappa(\phi^2) = \gamma(\phi^2) = 0 \cr \cr
a(p^2,\phi^2) = b(p^2,\phi^2) = c(p^2,\phi^2) = d(p^2,\phi^2)
=... =0 \cr \cr
\lambda(\phi^2) = \lambda = {\it scalar\ particle\ self\
coupling\ constant} \cr \cr m = {\it mass\ of\ the\ scalar\
particle} }
\end{equation}

\noindent and so the renormalized theory of quantum gravity
for a scalar field should have the form:

\begin{equation} L = -\sqrt{g}
\left( -2\Lambda + R +
\textstyle{1 \over 2} p^2 +
\textstyle{1 \over 2} m^2 \phi^2 +
\textstyle{1 \over 4!} \lambda \phi^4
\right)
\end{equation}

One might now worry about the renormalization group pulling
the coupling constants around. This is an open point to which
I feel one of several things might happen:

\begin{description} { \it

\item[$\bullet$] \hskip .59cm The couplings, set to zero at a
low energy scale, might reappear around the Plank scale.
Whether the resulting theory then makes sense is a matter for
dispute.

\item[$\bullet$] \hskip .59cm Certain coupling constants
(beyond those already set to zero) should be related, in order
that the beta functions of the zeroed couplings be zero (a
fixed point), so ensuring that all their couplings remain at
zero. This consistency condition could be the basis of a
unification scheme, although its implementation might not be
possible within the perturbative formulation. }
\end{description}

This is a highly non-trivial matter that needs looking at more
closely.

\subsection{Regularization Method}

On a diverse, but related track, one might wonder which
renormalization scheme to choose for implementing the scheme
proposed above. In this context analytic continuation [Bollini
et al., 1964; Speer, 1968; Salam and Strathdee, 1975; Dowker
and Critchley, 1976; Hawking, 1977] is very appealing in being
finite, and in this context there is an `unsung hero' in the
guise of operator regularization, which I think deserves a
mention [McKeon and Sherry, 1987; McKeon et al., 1987; McKeon
et al., 1988; Mann, 1988; Mann et al., 1989; Culumovic et al.,
1990; Shiekh, 1990].

In operator regularization one avoids the divergences by using
the analytic continuation:

\begin{equation}
\Omega^{-m} =
\lim \limits_{\varepsilon \to 0} {d^n \over d\varepsilon^n}
\left( {\varepsilon^n \over n! }
\Omega^{-\varepsilon - m}
\right)
\end{equation}

\noindent where $n$ is chosen large enough to eliminate the
infinities (the loop order is sufficient). Actually, operator
regularization is a bit of a misnomer, since it need not be
applied to an operator and does not just regulate, but also
renormalizes all in one. However, under this form of the
method {\it all} theories are finite and predictive (gravity
included). A little playing shows the above is simply an
automated system for minimal subtraction, and this realized,
the general form is easily located, and is given by:

\begin{equation}
\Omega^{-m} =
\lim \limits_{\varepsilon \to 0} {d^n \over d\varepsilon^n}
\left( (1+\alpha_1 \varepsilon +...+\alpha_n \varepsilon^n)
{\varepsilon^n \over n! }
\Omega^{-\varepsilon - m}
\right)
\end{equation}
\rightline{\small \it (the alphas being ambiguous)}

This form is not too powerful, and gravity must again be dealt
with as before, setting most of the final renormalized
parameters to zero on physical grounds.

That the earlier special form actually just locates and
zeros the divergences is perhaps most clearly seen by how it
treats some terms of the Maclaurin expansion. The taming of
$1$ yields:

$$\mathop {\lim}\limits_{\varepsilon \to 0}
{d \over {d\varepsilon }}\left( {\varepsilon .1} \right)=1$$

\noindent while, on the other hand, the taming of
${1 / \varepsilon}$ yields:

$$\mathop {\lim}\limits_{\varepsilon \to 0}
{d \over {d\varepsilon }}\left( {\varepsilon .{1 \over
\varepsilon }} \right)=0
$$

The method of operator regularization has the strength of
explicitly maintaining invariances, further even than
dimensional regularization, for dimension dependent
invariances are not disturbed. It is further blessed with the
feature of being finite throughout, as the Zeta function
technique [Salam and Strathdee, 1975; Dowker and Critchley,
1976; Hawking, 1975]. But unlike the Zeta function method, it
is not limited in applicability to one loop, being valid to
all orders.

To quickly see this method and its simplicity in action we
might look at a typical divergent one loop integral
intermediate result:

\begin{equation} I(p,m) =
\int_{-\infty}^{\infty} {{d^4x} \over {(2\pi)^4}}
\int_{0}^{1} dx {{p^2l^2+2m^2p.l-2m^4}
\over {\left[ l^2+m^2x+p^2(1-x)+2p.l(1-x) \right]^2}}
\end{equation}

\noindent which we tame with:

\begin{equation}
\Omega^{-2} =
\lim \limits_{\varepsilon \to 0} {d \over d\varepsilon}
\left( (1+\alpha \varepsilon) \varepsilon
\Omega^{-\varepsilon - 2}
\right)
\end{equation}

\noindent to yield the finite object:

\begin{equation} I(p,m) =
\int_{0}^{1} dx
\lim \limits_{\varepsilon \to 0} {d \over d\varepsilon}
\int_{-\infty}^{\infty} {{d^4x} \over {(2\pi)^4}}
\left(
\varepsilon (1+\alpha \varepsilon) {{p^2l^2+2m^2p.l-2m^4}
\over {\left[ l^2+m^2x+p^2(1-x)+2p.l(1-x) \right]^
{\varepsilon+2}}}
\right)
\end{equation}

\noindent from which we proceed; to yield:

\begin{equation} {m^4 \over (4\pi)^2}
\left(
\left( 3 + 2{p^2 \over m^2} + {m^2 \over p^2}
\right)
\ln ( 1 + {p^2 \over m^2} ) - 1 - {5 \over 2} {p^2 \over m^2}
- {1 \over 6} {p^4 \over m^4} + 2 \left( 1 + {p^2 / m^2}
\right)
\left( \ln ({m^2 / \mu^2} ) - \alpha \right)
\right)
\end{equation}

\noindent which actually has no divergence at $p=0$. The
factor $\mu$ which leads to the renormalization group appears
on dimensional grounds. As is typical at one loop (but not
beyond), the arbitrary factor $\alpha$ can be reabsorbed into
the parameter $\mu$.

\subsection{Discussion}

We are now left with a finite theory that has few arbitrary
constants, and so is predictive. Despite the present lack of
experimental data to test it against, and regardless of the
patch work line of reasoning invoked to arrive at this
hypothesis, one might alter perspective and simply be
interested in investigating the consequences of such a scheme
for its own sake, where many of the arbitrary factors have
been set to zero, for whatever reason. At this stage any well
behaved, finite theory, is worth investigating; and it is
unfortunate that we don't have the guiding hand of mother
nature to assist us in this guessing game.\\

{\Large \bf Acknowledgments}

For having invited me to participate in such a vibrant conference,
despite being the carrier of neutral tidings,
my gratitude goes to the director Michael Holzscheiter.

The beautiful location went a long way toward making us reflect
at length at the the very nature we came together to investigate.

I should also like to thank John Strathdee and Seifallah Randjbar
Daemi for listening to and commenting upon, if not necessarily
agreeing with, my ideas. Special thanks are extended to Gerry
McKeon and Roberto Percacci for a responsive ear and
constructive criticism. \\

{\Large \bf References}

\begin{description} {\small

\item[$\bullet$] \hskip .59cm {\bf A. Shiekh},
{\it `The Perturbative Quantization of Gravity'}, in
``Problems on High Energy Physics and Field Theory'',
Proceedings of the XVII workshop 1994, pp. 156-165,
Protvino, 1995.
\\
{\bf A. Shiekh},
{\it `Quantizing Orthodox Gravity'}, Can. J. Phys.,
{\bf 74}, 1996, 172-.

\item[$\bullet$] \hskip .59cm {\bf C.J. Isham}, {\it `Quantum
Gravity - An Overview'}, in ``Quantum Gravity 2: A Second
Oxford Symposium'', pp. 1-62, eds. C.J. Isham, R. Penrose and
D.W. Sciama, (Oxford University Press, Oxford, 1981).

\item[$\bullet$] \hskip .59cm {\bf R.P. Feynman}, {\it
``Lectures on Gravitation''} (Caltech, 1962-1963, unpublished)
\\ {\bf T.W.B. Kibble}, {\it `Is a
Semi-Classical Theory of Gravity Viable?'}, in ``Quantum
Gravity 2: A Second Oxford Symposium'', pp. 63-80, eds. C.J.
Isham, R. Penrose and D.W. Sciama, (Oxford University Press,
Oxford, 1981).

\item[$\bullet$] \hskip .59cm {\bf P. van Nieuwenhuizen},
{\it `Radiation of Massive Gravitation'}, Phys. Rev.,
{\bf D7}, 1973, 2300-.

\item[$\bullet$] \hskip .59cm {\bf P. Ramond}, {\it ``Field
Theory: A Modern Primer''},
 2nd Ed (Addison-Wesley, 1990).
\\ {\bf J. Collins}, {\it ``Renormalization''}, (Cambridge
University Press, London, 1984).

\item[$\bullet$] \hskip .59cm {\bf M.J.G. Veltman}, {\it
`Quantum Theory of Gravitation'}, Les Houches XXVIII,
``Methods In Field Theory'', pp. 265-327, eds. R. Ballan and
J. Zinn-Justin, (North-Holland, Amsterdam, 1976).

\item[$\bullet$] \hskip .59cm {\bf C. Bollini, J. Giambiagi
and A. Dom\`{\i}nguez}, {\it `Analytic Regularization and the
Divergences of Quantum Field Theories'}, Nuovo Cimento {\bf
31}, 1964, 550-.
\\ {\bf E. Speer}, {\it `Analytic Renormalization'}, J. Math.
Phys. {\bf 9}, 1968, 1404-.
\\ {\bf A. Salam and J. Strathdee}, {\it `Transition
Electromagnetic Fields in Particle Physics'}, Nucl. Phys. {\bf
B90}, 1975, 203-.
\\ {\bf J. Dowker and R. Critchley}, {\it `Effective Lagrangian
and energy-momentum tensor in de sitter space'}, Phys. Rev.,
{\bf D13}, 1976, 3224-.
\\ {\bf S. Hawking}, {\it `Zeta Function Regularization of
Path Integrals in Curved Space'}, Commun. Math. Phys., {\bf
55}, 1977, 133-.

\item[$\bullet$] \hskip .59cm {\bf D. McKeon and T. Sherry},
{\it `Operator Regularization of Green's Functions'}, Phys.
Rev. Lett., {\bf 59}, 1987, 532-.
\\ {\bf D. McKeon and T. Sherry}, {\it `Operator
Regularization and one-loop Green's functions'}, Phys. Rev.,
{\bf D35}, 1987, 3854-.
\\ {\bf D. McKeon, S. Rajpoot and T. Sherry}, {\it `Operator
Regularization with Superfields'}, Phys. Rev., {\bf D35},
1987, 3873-.
\\ {\bf D. McKeon, S. Samant and T. Sherry}, {\it `Operator
regularization beyond lowest order'}, Can. J. Phys., {\bf 66},
1988, 268-.
\\ {\bf R. Mann.}, {\it `Zeta function regularization of
Quantum Gravity'}, In Proceedings of the cap-nserc Summer
Workshop on Field Theory and Critical Phenomena. Edited by G.
Kunstatter, H. Lee, F. Khanna and H. Limezawa, World
Scientific Pub. Co. Ltd., Singapore, 1988, p. 17-.
\\ {\bf R. Mann, D. McKeon, T. Steele and T. Tarasov}, {\it
`Operator Regularization and Quantum Gravity'}, Nucl. Phys.,
{\bf B311}, 1989, 630-.
\\ {\bf L. Culumovic, M. Leblanc, R. Mann, D. McKeon and T.
Sherry}, {\it `Operator regularization and multiloop Green's
functions'}, Phys. Rev., {\bf D41}, 1990, 514-.
\\ {\bf A. Shiekh}, {\it `Zeta-function regularization of
quantum field theory'}, Can. J. Phys., {\bf 68}, 1990, 620-.

}
\end{description}

\end{document}